\documentclass[11pt,a4paper]{article}
\usepackage[utf8]{inputenc}
\usepackage{graphicx}
\usepackage{braket}
\usepackage{tikz}
\usepackage{tikz-3dplot}
\usepackage{quantikz}
\usetikzlibrary{decorations.pathreplacing}
\usepackage{booktabs}
\usepackage{amsmath,amssymb,mathtools}
\usepackage{hyperref}
\hypersetup{colorlinks=true,linkcolor=blue,citecolor=blue,urlcolor=blue}
\usepackage{geometry}
\date{}
\begin{document}
\title{{\bf Qubit representation of the mass observable in quantized Schwarzschild interior}}

\author{Babak Vakili\thanks{email:
ba.vakili@iau.ac.ir}\\\\{\small {\it
Department of Physics, CT.C., Islamic Azad
University, Tehran, Iran}}} \maketitle

\begin{abstract}
We develop a quantum-computational framework for the mass observable
of the Wheeler--DeWitt quantized Schwarzschild black-hole interior.
Working in the physical sector spanned by correlated two-mode
oscillator states, we construct finite-dimensional representations of
the mass operator and map them onto qubit operators through Pauli
decomposition. We analyze the resulting representations for increasing
truncation sizes and establish analytically the scaling of the number
of nonvanishing Pauli terms and independent measurement settings.
These results provide a quantitative characterization of the quantum
resources required to encode and measure the black-hole mass
observable. As a representative benchmark, we implement a
two-qubit variational quantum eigensolver for the smallest nontrivial
truncation and find excellent agreement with the exact finite-dimensional
spectrum. We further describe a quantum-circuit procedure for
verifying the Wheeler--DeWitt constraint independently of the mass
estimation. Our results provide a direct connection between
Wheeler--DeWitt quantum gravity and quantum computation and establish
a structured route toward the quantum simulation of gravitational
observables.
\vspace{5mm}\noindent\\
Keywords: Quantum black holes; Quantum computation; Quantum complexity; Pauli decomposition; Variational quantum eigensolver
\end{abstract}

\section{Introduction}
\label{sec:introduction}

Black holes provide one of the most intriguing arenas in which the
fundamental principles of general relativity, quantum mechanics, and
thermodynamics meet. The Schwarzschild solution of Einstein's field
equations represents the simplest black-hole geometry and provides a
natural starting point for investigating the quantum structure of
black-hole spacetimes~\cite{Schwarzschild1916}. The discovery of black
hole thermodynamics and Hawking radiation further demonstrated that
black holes possess entropy and temperature, establishing a deep
connection between gravity, quantum theory, and information
~\cite{Bekenstein1973,Hawking1975,Hawking1976}. Understanding the
quantum dynamics of black holes therefore remains an important problem
in the search for a consistent description of gravity at the quantum
level. In particular, the classical singularity appearing in the
interior of the Schwarzschild geometry signals a regime in which the
classical description of spacetime is expected to break down, thereby
providing a natural setting for quantum-gravitational investigations.

Canonical quantization provides one of the earliest and most direct
approaches to quantum cosmology and quantum gravity. In this framework,
the Wheeler--DeWitt (WDW) equation replaces the classical Hamiltonian
constraint by an operator equation acting on the wave function of the
gravitational system~\cite{DeWitt1967}. Applying this procedure to
symmetry-reduced black-hole geometries leads to finite-dimensional
minisuperspace models in which the gravitational degrees of freedom
can be treated as quantum mechanical variables. Such models provide
tractable settings in which questions concerning the quantum states,
spectra of physical observables, and the fate of the classical
singularity can be investigated without requiring a complete
formulation of quantum gravity. WDW quantization has
indeed been extensively applied to black-hole interiors and
gravitational collapse, where quantum effects may modify the classical
singularity structure~\cite{Kiefer2007,Modesto2004}.

At the same time, quantum computing has emerged as a new computational
paradigm based on the coherent manipulation of quantum states. The
concept of using quantum mechanical systems to perform computation
dates back to the foundational proposals of quantum computation
~\cite{Benioff1980,Feynman1982,Deutsch1985}, while the discovery of
quantum algorithms demonstrated that quantum
computers can provide computational capabilities fundamentally
different from those of classical machines~\cite{Shor1994}. More
generally, quantum simulation was identified as one of the most
promising applications of quantum computers, since a quantum processor
can naturally encode and manipulate quantum states whose classical
description may require exponentially large resources
~\cite{Lloyd1996,Zalka1998,Preskill1998}. The rapid development of
quantum hardware and variational algorithms has subsequently made the
study of physically motivated quantum systems on near-term quantum
processors an active area of research.

The relevance of quantum computation extends naturally to fundamental
physics. Quantum simulation of quantum field theories, strongly
correlated systems, and gauge theories has become an important
research direction, with quantum computers offering a possible route
toward studying regimes in which classical numerical methods become
particularly challenging~\cite{Preskill2018}.
Since gravity itself is fundamentally quantum in any complete
description of nature, quantum information and quantum computation
have also been proposed as useful languages and computational tools
for investigating quantum-gravitational systems. Early ideas relating
quantum information and quantum gravity include the proposal of
describing quantum-gravitational degrees of freedom in terms of
quantum bits and the connection between holography, quantum geometry,
and black-hole entropy~\cite{Zizzi2000,Zizzi2001}. More recently, the
controlled manipulation of massive quantum systems has opened new
possibilities for experimentally probing the interface between
quantum mechanics and gravity, including gravity-mediated
entanglement and quantum tests of gravitational interactions \cite{Bose2025}. These developments reinforce the broader
perspective that quantum information and quantum computation may
provide useful tools for studying gravitational systems rather than
serving solely as technologies for conventional computational
problems.

A particularly direct realization of this idea is provided by
quantum-computational studies of minisuperspace black-hole models.
Joseph \textit{et al.} demonstrated that the WDW
Hamiltonian and mass operators of Schwarzschild--de Sitter
black holes and Kantowski--Sachs cosmology can be represented by
finite-dimensional matrices and subsequently implemented on quantum
computers using the variational quantum eigensolver (VQE)
~\cite{Joseph2022}. Their results showed that the low-dimensional
case can be treated with high accuracy, while the increasing size of
the Hilbert space introduces additional challenges for the variational
ansatz. This work provides an important motivation for investigating
more systematically how gravitational observables can be encoded into
qubits and what quantum resources are required for their measurement \cite{Chandra}. Also, a related information-theoretic approach was recently proposed
by Nakarachinda {\it et al.}, who modelled the Schwarzschild--de Sitter
black hole as a correlated two-qubit system by associating its
black-hole and cosmological horizons with two subsystems and
identifying their entropies with those of the corresponding
qubit states \cite{Nakarachinda2025}. Their approach employs the quantum-information concepts in the study of black-hole
physics, so their objectives and interpretations are somehow different from us.
In this paper, qubits encode a truncated Hilbert space
in which the mass operator and the WDW constraint
can be investigated using quantum-computational methods.
In contrast, the correlated-qubit description of \cite{Nakarachinda2025}
focuses on the entropic properties and
correlations of the two horizons. The two frameworks may
therefore be viewed as complementary rather than equivalent.

In the present work, we focus on the mass observable of the quantized Schwarzschild black-hole interior. Starting
from the physical states of the minisuperspace model, we construct a
finite-dimensional truncation of the mass operator in the basis of
physical two-mode states $|n,n\rangle$. We then translate the resulting
matrix representation into a sum of tensor products of Pauli
operators, thereby obtaining an explicit qubit representation of the
black-hole mass observable. Beyond the construction of individual
finite-dimensional representations, we investigate their scaling with
the truncation dimension $N$. In particular, we establish analytically
that the number of nonvanishing Pauli strings grows as
$2N-1$, while the number of qubits and independent measurement settings
exhibit logarithmic scaling with $N$.

We further use the two-qubit $N=4$ representation as a representative
variational benchmark. A shallow VQE circuit is employed to estimate
the lowest eigenvalue of the truncated mass operator, and the
variational result is compared with exact diagonalization. Finally,
we discuss an independent quantum-circuit procedure for verifying the
WDW constraint through measurements of the physical
correlation between the two oscillator registers. In this way, the
present work connects the canonical quantization of a black-hole
interior with an explicit matrix-to-Pauli-to-qubit construction and
provides a quantitative characterization of the quantum resources
required to encode and measure a physical gravitational observable.

\section{Classical Schwarzschild minisuperspace}

We begin with the Schwarzschild spacetime, whose metric in the
standard Schwarzschild coordinates is given by

\begin{equation}
ds^2=-f(r)\,dt^2+\frac{dr^2}{f(r)}+r^2d\Omega^2,\hspace{5mm}f(r)=1-\frac{2GM}{r}.
\label{eq:Schwarzschild_metric}
\end{equation}
The Schwarzschild horizon is located at $r_h=2GM$, and our interest is in the quantum description of the black-hole interior,
corresponding to the region $0<r<r_h$. In this region, $f(r)<0$, and consequently the causal character of the
coordinates $t$ and $r$ is interchanged. In particular, the coordinate
$r$ becomes timelike, whereas $t$ becomes spacelike. Introducing an interior time coordinate $\tau$ through $\tau=r$, and denoting the homogeneous spatial coordinate associated with the
Schwarzschild time coordinate by $x$, the interior metric can be written as

\begin{equation}
ds^2=-\frac{d\tau^2}{\displaystyle\frac{2GM}{\tau}-1}+\left(\frac{2GM}{\tau}-1\right)dx^2+\tau^2d\Omega^2.
\label{eq:Schwarzschild_interior}
\end{equation}
Therefore, the metric coefficients are promoted to dynamical
variables depending on a single evolution parameter $t$. We therefore
write the minisuperspace variables as $a=a(t)$ and $b=b(t)$, and begin with the minisuperspace description of the Schwarzschild
black hole. The resulting reduced metric has the Kantowski--Sachs form

\begin{equation}
ds^2=-N^2(t)dt^2+a^2(t)dx^2+b^2(t)d\Omega^2,
\label{eq:KS_minisuperspace_metric}
\end{equation}
where $N(t)$ is the lapse function. The coordinate $x$ is homogeneous, while the two functions $a(t)$
and $b(t)$ contain the remaining gravitational degrees of freedom. The use of the parameter $t$ in the reduced description should not be
confused with the Schwarzschild time coordinate appearing in
Eq.~\eqref{eq:Schwarzschild_metric}. In particular, the dots in the
minisuperspace Lagrangian denote derivatives with respect to the
chosen evolution parameter $t$ of the reduced system.

The classical Schwarzschild interior is recovered as a particular
solution of the reduced equations of motion. The minisuperspace
description therefore provides a finite-dimensional dynamical system
whose canonical quantization can be used to investigate the quantum
properties of the black-hole interior. So, for the Schwarzschild geometry, the corresponding reduced Lagrangian is

\begin{equation}
L=-\frac{b\dot{b}\dot{a}}{N}-\frac{a\dot{b}^{\,2}}{2N}+\frac{Na}{2}.
\label{eq:Schwarzschild_Lagrangian}
\end{equation}
The canonical momenta conjugate to $a$ and $b$ are defined by

\begin{equation}
p_a=\frac{\partial L}{\partial\dot{a}}=-\frac{b\dot{b}}{N},
\qquad
p_b=\frac{\partial L}{\partial\dot{b}}=-\frac{b\dot{a}}{N}
-\frac{a\dot{b}}{N}.
\end{equation}
The canonical Hamiltonian is obtained through the Legendre
transformation,

\begin{equation}
H_{\mathrm{can}}=p_a\dot{a}+p_b\dot{b}-L,
\end{equation}
which yields

\begin{equation}
H_{\mathrm{can}}=N\left[\frac{a}{2b^2}p_a^2-\frac{1}{b}p_ap_b-\frac{a}{2}\right].
\label{eq:Schwarzschild_Hcan}
\end{equation}
Since the lapse function $N$ is a Lagrange multiplier, variation of
the action with respect to $N$ yields the Hamiltonian constraint

\begin{equation}
H=-\frac{a}{2b^2}p_a^2+\frac{1}{b}p_ap_b+\frac{a}{2}=0.
\label{eq:Schwarzschild_H}
\end{equation}
The mass function of the Schwarzschild black hole can be expressed
in terms of the minisuperspace variables and their conjugate momentum. For a spherically symmetric spacetime, the Misner--Sharp mass is given
by \cite{MisnerSharp1964,Hayward1996}

\begin{equation}
M=\frac{b}{2}\left(1-g^{\mu\nu}\partial_\mu b\,\partial_\nu b\right),
\label{eq:misner_sharp_mass}
\end{equation}
where $b$ is the areal radius.  For the Kantowski--Sachs form (\ref{eq:KS_minisuperspace_metric}) of the
Schwarzschild interior metric this relation becomes

\begin{equation}
M=\frac{b}{2}\left(1+\frac{\dot b^2}{N^2}\right),
\label{eq:mass_velocity_form}
\end{equation}
which in terms of the canonical momenta takes the form

\begin{equation}
M=
\frac{p_a^2}{2b}
+\frac{b}{2}.
\label{eq:Schwarzschild_M}
\end{equation}
At the classical level, the Hamiltonian constraint and the mass
function satisfy
\begin{equation}
\{H,M\}=0.
\label{eq:Schwarzschild_Poisson}
\end{equation}
Therefore, the mass is conserved along the Hamiltonian flow generated
by the constraint. The classical problem is consequently characterized by the two
relations
\begin{equation}
H=0,
\qquad
M=\mathrm{constant}.
\label{eq:Schwarzschild_classical_relations}
\end{equation}
These equations will provide the starting point for the canonical
transformation and subsequent quantization of the Schwarzschild
minisuperspace.

We now introduce a canonical transformation that simplifies the
structure of the Schwarzschild minisuperspace Hamiltonian. We define
the new configuration variables by the canonical transformation

\begin{eqnarray}
\left\{
\begin{array}{ll}
u=\sqrt{b}\,(a+1),\\\\
v=\sqrt{b}\,(a-1),\\\\
p_u=-\frac{2v}{(u-v)^2}p_a+\frac{u-v}{2}p_b,\\\\
p_v=\frac{2u}{(u-v)^2}p_a-\frac{u-v}{2}p_b,
\end{array}
\right.
\end{eqnarray}
the inverse of which takes the form

\begin{eqnarray}
\left\{
\begin{array}{ll}
a=\frac{u+v}{u-v},\\\\
b=\frac{(u-v)^2}{4},\\\\
p_a=\frac{u-v}{2}\,(p_u+p_v),\\\\
p_b=2\frac{up_u+vp_v}{(u-v)^2}.
\end{array}
\right.
\end{eqnarray}
Since the transformation defined above is canonical by construction, it preserves the Poisson brackets,

\begin{equation}
\{u,p_u\}=1,
\qquad
\{v,p_v\}=1,
\end{equation}
with all other elementary Poisson brackets vanishing. In terms of these new variables the Hamiltonian constraint and the Schwarzschild mass
function take the form

\begin{equation}
2bH=p_u^2-p_v^2+\frac{1}{4}(u^2-v^2),
\label{eq:scaled_H_uv}
\end{equation}
and 

\begin{equation}
4M=2(p_u+p_v)^2+\frac{1}{2}(u-v)^2.
\label{eq:4M_uv}
\end{equation}
Because $b>0$ in the physical region, the constraints $H=0$ and $2bH=0$,
are classically equivalent. The latter form is considerably simpler
and will be useful for the subsequent quantization. Equations~\eqref{eq:scaled_H_uv} and
\eqref{eq:4M_uv} constitute the Hamiltonian constraint and the mass
function obtained directly from the original Schwarzschild
minisuperspace Lagrangian and the canonical transformation
defined above. A closely related application of such oscillator
framework was also developed in our earlier work on black-hole
evaporation and information flow \cite{BayenatVakili2026}. In that study, a two-mode model was constructed in which
an effective geometric degree of freedom and a representative
Hawking-radiation mode are described by coupled oscillators with opposite signs in their free Hamiltonians.

\section{Canonical quantization and the WDW equation}

The Hamiltonian constraint obtained from the classical minisuperspace Lagrangian is

\begin{equation}
\mathcal{H}=p_u^2-p_v^2+\frac{1}{4}(u^2-v^2)=0.
\label{eq:classical_constraint_uv}
\end{equation}
We now perform canonical quantization by promoting the momenta to differential operators. Consequently, the Hamiltonian constraint becomes the
WDW equation \footnote{The indefinite-oscillator structure of the WDW
equation also connects the present analysis to earlier work
on exact wave packets in quantum cosmology. In particular, the wave-packet solutions for minisuperspace
models involving a Friedmann universe with a conformally
coupled scalar field, both with and without a cosmological
constant are constructed in \cite{Kiefer1990}. This work provides a useful
mathematical context for the study of wave packets in
indefinite-oscillator models. However, the physical
minisuperspace model considered there differs from the
Schwarzschild-interior model studied here. The relevance
of the exact solutions to the present system therefore
depends on establishing an explicit correspondence between
the respective WDW equations and their boundary
conditions. Our main objective is
instead to construct a finite-dimensional representation
of the mass operator and investigate its implementation
in terms of qubits.}

\begin{equation}
\left[-\frac{\partial^2}{\partial u^2}+\frac{\partial^2}{\partial v^2}+\frac{1}{4}(u^2-v^2)\right]\Psi(u,v)=0.
\label{eq:WDW_uv}
\end{equation}
This equation has the structure of an oscillator--ghost-oscillator system and is separable. We therefore
seek solutions of the form

\begin{equation}
\Psi(u,v)=\psi_u(u)\psi_v(v),
\label{eq:separable_wavefunction}
\end{equation}
substitution of which into Eq.~\eqref{eq:WDW_uv} gives a pair of ordinary differential equations

\begin{equation}
-\frac{d^2\psi_u}{du^2}+\frac{u^2}{4}\psi_u=\epsilon\psi_u,\hspace{5mm}-\frac{d^2\psi_v}{dv^2}+\frac{v^2}{4}\psi_v=\epsilon\psi_v,
\label{eq:u_oscillator}
\end{equation}
where $\epsilon$ is a separation constant. Both equations have the form of a one-dimensional harmonic
oscillator. Indeed, introducing $u=\sqrt{2}\,x$, equations ~\eqref{eq:u_oscillator} become

\begin{equation}
\left[-\frac{1}{2}\frac{d^2}{dx^2}+\frac{x^2}{2}\right]\psi_u=\epsilon\psi_u, \hspace{5mm} \left[-\frac{1}{2}\frac{d^2}{dx^2}+\frac{x^2}{2}\right]\psi_v=\epsilon\psi_v.
\end{equation}
Hence, the normalizable eigenvalues are

\begin{equation}
\epsilon_{n_u}=n_u+\frac{1}{2},\hspace{5mm}\epsilon_{n_v}=n_v+\frac{1}{2},
\qquad
n_u,n_v=0,1,2,\ldots.
\label{eq:epsilon_u}
\end{equation}
The WDW constraint requires the two separation constants to be equal. Therefore,

\begin{equation}
\epsilon_{n_u}=\epsilon_{n_v},\Rightarrow n_u=n_v.
\label{eq:physical_quantum_numbers}
\end{equation}
The physical Schwarzschild quantum states can consequently be written
as
\begin{equation}
\Psi_n(u,v)
=
\psi_n(u)\psi_n(v),
\qquad
n=0,1,2,\ldots.
\label{eq:physical_WDW_states}
\end{equation}
where the corresponding normalized eigenfunctions, expressed directly in
terms of the original variables $u$ and $v$ are
\begin{equation}
\psi_n(u)=\frac{1}{\pi^{1/4}\sqrt{2^n n!}}e^{-\frac{u^2}{4}}H_n\left(\frac{u}{\sqrt{2}}\right),\hspace{5mm}\psi_n(v)=\frac{1}{\pi^{1/4}\sqrt{2^n n!}}e^{-\frac{v^2}{4}}H_n\left(\frac{v}{\sqrt{2}}\right).
\label{eq:normalized_u_wavefunction}
\end{equation}
where $H_n$ denotes the Hermite polynomial of degree $n$. Therefore, the eigenstates of the WDW equation are explicitly given by

\begin{equation}
\Psi_n(u,v)=\frac{1}{\sqrt{\pi}\,2^n n!}\exp\left[-\frac{u^2+v^2}{4}\right]H_n\left(\frac{u}{\sqrt{2}}\right)H_n\left(\frac{v}{\sqrt{2}}\right).
\label{eq:explicit_physical_states}
\end{equation}
A general physical state can consequently be expressed as a
superposition of the basis states,

\begin{equation}
\Psi(u,v)=\sum_{n=0}^{\infty}c_n\Psi_n(u,v),
\label{eq:general_physical_state}
\end{equation}
where the coefficients satisfy

\begin{equation}
\sum_{n=0}^{\infty}|c_n|^2=1.
\label{eq:state_normalization}
\end{equation}
The mass operator obtained from the classical Schwarzschild mass
function (\ref{eq:4M_uv}) is

\begin{equation}
4\hat{M}=2(\hat{h}_u+\hat{h}_v)+4\hat{p}_u\hat{p}_v-\hat{u}\hat{v}, 
\label{eq:mass_operator_oscillator_form}
\end{equation}
in which we introduced the oscillator Hamiltonians

\begin{equation}
\hat{h}_u=\hat{p}_u^2+\frac{1}{4}\hat{u}^2,
\qquad
\hat{h}_v=\hat{p}_v^2+\frac{1}{4}\hat{v}^2.
\label{eq:oscillator_hamiltonians}
\end{equation}
Since

\begin{equation}
\hat{h}_u\ket{n_u,n_v}=\left(n_u+\frac{1}{2}\right)\ket{n_u,n_v},\hspace{5mm}\hat{h}_v\ket{n_u,n_v}=\left(n_v+\frac{1}{2}\right)\ket{n_u,n_v},
\end{equation}
the diagonal contribution of the mass operator in the physical
subspace is

\begin{equation}
\bra{n,n}\hat{M}\ket{n,n}=n+\frac{1}{2}.
\label{eq:mass_diagonal_element}
\end{equation}
The remaining terms, $4\hat{p}_u\hat{p}_v-\hat{u}\hat{v}$,
couple neighboring oscillator states. In particular, since both
$\hat{u}$ and $\hat{p}_u$ change the oscillator quantum number by
one unit, and similarly for the $v$ degree of freedom, these terms
connect

\begin{equation}
\ket{n,n}
\longleftrightarrow
\ket{n+1,n+1}
\end{equation}
and

\begin{equation}
\ket{n,n}
\longleftrightarrow
\ket{n-1,n-1}.
\end{equation}
Consequently, the mass operator preserves the physical subspace
defined by the WDW constraint and has a tridiagonal
representation in the basis

\begin{equation}
\left\{
\ket{0,0},
\ket{1,1},
\ket{2,2},
\ldots
\right\}.
\label{eq:physical_mass_basis}
\end{equation}

\subsection{Matrix representation of the mass operator}

To determine the action of the mass operator on the physical
WDW states, we introduce the ladder operators associated
with the two oscillator degrees of freedom. For the $u$ mode, we define

\begin{equation}
\hat{u}=\hat{a}_u+\hat{a}_u^\dagger,
\qquad
\hat{p}_u=\frac{i}{2}\left(\hat{a}_u^\dagger-\hat{a}_u\right),
\label{eq:u_ladder_operators}
\end{equation}
while for the $v$ mode,

\begin{equation}
\hat{v}=\hat{a}_v+\hat{a}_v^\dagger,
\qquad
\hat{p}_v=\frac{i}{2}\left(\hat{a}_v^\dagger-\hat{a}_v\right).
\label{eq:v_ladder_operators}
\end{equation}
The ladder operators satisfy
\begin{equation}
[\hat{a}_u,\hat{a}_u^\dagger]=1,
\qquad
[\hat{a}_v,\hat{a}_v^\dagger]=1,
\end{equation}
with the operators belonging to different modes commuting with each
other. The oscillator Hamiltonians introduced previously can then be written
as

\begin{equation}
\hat{h}_u=\hat{a}_u^\dagger\hat{a}_u+\frac{1}{2},
\qquad
\hat{h}_v=\hat{a}_v^\dagger\hat{a}_v+\frac{1}{2}.
\label{eq:oscillator_number_operators}
\end{equation}
Also, the mass operator (\ref{eq:mass_operator_oscillator_form}) takes the form

\begin{equation}
\hat{M}=\frac{1}{2}\left(\hat{n}_u+\hat{n}_v+1\right)-\frac{1}{2}\left(\hat{a}_u\hat{a}_v+\hat{a}_u^\dagger\hat{a}_v^\dagger\right),
\label{eq:mass_operator_ladder}
\end{equation}
where

\begin{equation}
\hat{n}_u=\hat{a}_u^\dagger\hat{a}_u,
\qquad
\hat{n}_v=\hat{a}_v^\dagger\hat{a}_v.
\end{equation}
The physical WDW states satisfy $\ket{n}_{\rm phys}\equiv \ket{n,n}$. Using

\begin{equation}
\hat{a}_u\hat{a}_v\ket{n,n}=n\ket{n-1,n-1},
\end{equation}
and

\begin{equation}
\hat{a}_u^\dagger\hat{a}_v^\dagger\ket{n,n}=(n+1)\ket{n+1,n+1},
\end{equation}
we obtain

\begin{equation}
\hat{M}\ket{n,n}=\left(n+\frac{1}{2}\right)\ket{n,n}-\frac{n}{2}\ket{n-1,n-1}-\frac{n+1}{2}\ket{n+1,n+1}.
\label{eq:mass_action_physical_states}
\end{equation}
Therefore, the matrix elements of the mass operator in the physical
basis are

\begin{equation}
M_{mn}=\left(n+\frac{1}{2}\right)\delta_{mn}-\frac{n}{2}\delta_{m,n-1}-\frac{n+1}{2}\delta_{m,n+1}.
\label{eq:mass_matrix_elements}
\end{equation}
The resulting matrix has a tridiagonal structure,

\begin{equation}
\hat{M}=\begin{pmatrix}
\frac{1}{2} & -\frac{1}{2} & 0 & 0 & \cdots\\
-\frac{1}{2} & \frac{3}{2} & -1 & 0 & \cdots\\
0 & -1 & \frac{5}{2} & -\frac{3}{2} & \cdots\\
0 & 0 & -\frac{3}{2} & \frac{7}{2} & \cdots\\
\vdots & \vdots & \vdots & \vdots & \ddots
\end{pmatrix}.
\label{eq:mass_tridiagonal_matrix}
\end{equation}
It is important to emphasize that the diagonal quantity
$n+1/2$ is not, by itself, a mass eigenvalue, since
$\ket{n,n}$ is not an eigenstate of $\hat{M}$. The physical mass
spectrum must instead be obtained by diagonalizing the matrix
representation of $\hat{M}$ in the physical subspace.

\subsection{Spectral analysis of the mass operator}

The spectral problem of the mass operator can be solved analytically.
Let a generalized eigenstate of $\hat{M}$ be written in the physical
basis as

\begin{equation}
\ket{m}=\sum_{n=0}^{\infty}c_n(m)\ket{n,n},
\label{eq:mass_generalized_eigenstate}
\end{equation}
where

\begin{equation}
\hat{M}\ket{m}=m\ket{m}.
\label{eq:mass_eigenvalue_equation}
\end{equation}
Using Eq.~\eqref{eq:mass_action_physical_states}, the coefficients
$c_n(m)$ satisfy the recurrence relation

\begin{equation}
(n+1)c_{n+1}(m)+n c_{n-1}(m)-(2n+1-2m)c_n(m)=0.
\label{eq:mass_recurrence_laguerre}
\end{equation}
This recurrence relation coincides with the standard recurrence
relation for the Laguerre polynomials \cite{Abra},

\begin{equation}
(n+1)L_{n+1}(x)+nL_{n-1}(x)-(2n+1-x)L_n(x)=0,
\label{eq:laguerre_recurrence}
\end{equation}
provided that $x=2m$. Consequently, the generalized eigenstate coefficients are

\begin{equation}
c_n(m)=L_n(2m).
\label{eq:laguerre_coefficients}
\end{equation}
The corresponding generalized eigenstates are therefore

\begin{equation}
\ket{m}=\sum_{n=0}^{\infty}L_n(2m)\ket{n,n}.
\label{eq:generalized_mass_eigenstates}
\end{equation}
Since the Laguerre polynomials satisfy \cite{Abra}

\begin{equation}
\int_0^\infty e^{-x}L_n(x)L_k(x)\,dx=\delta_{nk},
\label{eq:laguerre_orthogonality}
\end{equation}
the transformation from the physical oscillator basis to the
spectral representation of $\hat{M}$ is characterized by the measure

\begin{equation}
d\mu(m)=2e^{-2m}\,dm.
\label{eq:mass_spectral_measure}
\end{equation}
Accordingly, the generalized mass eigenstates may be normalized according to

\begin{equation}
\langle m|m'\rangle=\frac{\delta(m-m')}{2e^{-2m}},
\end{equation}
such that the resolution of the identity takes the form

\begin{equation}
\int_0^\infty
2e^{-2m}dm\,|m\rangle\langle m|=\mathbb I.
\end{equation}
Also, the physical Hilbert space admits the spectral representation

\begin{equation}
\mathcal{H}_{\rm phys}
\simeq L^2\left([0,\infty),2e^{-2m}dm\right),
\label{eq:spectral_hilbert_space}
\end{equation}
in which the mass operator acts by multiplication

\begin{equation}
(\hat{M}\phi)(m)=m\phi(m),
\label{eq:mass_multiplication_operator}
\end{equation}
and a normalizable physical state can takes the form

\begin{equation}
|\Psi\rangle=\int_0^\infty 2e^{-2m}dm\,A(m)|m\rangle,
\end{equation}
where the spectral amplitude satisfies

\begin{equation}
\int_0^\infty 2e^{-2m}|A(m)|^2dm=1.
\end{equation}
Using the expansion of the generalized eigenstates in the
oscillator basis, the physical state can equivalently be written as

\begin{equation}
|\Psi\rangle=\sum_{n=0}^{\infty}c_n|n,n\rangle,
\label{Psi}
\end{equation}
where

\begin{equation}
c_n=\int_0^\infty 2e^{-2m}dm\,A(m)L_n(2m).
\end{equation}
The orthogonality and completeness of the Laguerre basis ensure
that $\sum_{n=0}^{\infty}|c_n|^2=1$. For concreteness, one may choose a Gaussian spectral wave packet

\begin{equation}
A(m)=\mathcal N_A\exp\left[-\frac{(m-m_0)^2}{4\sigma_m^2}\right],\qquad m\geq0,
\end{equation}
where the normalization constant is determined from

\begin{equation}
\mathcal N_A^{-2}=\int_0^\infty2e^{-2m}\exp\left[-\frac{(m-m_0)^2}{2\sigma_m^2}\right]dm.
\end{equation}
The resulting coefficients therefore form a two-parameter family,

\begin{equation}
c_n=c_n(m_0,\sigma_m),
\end{equation}
rather than an arbitrary set of independent amplitudes.

In summary, the above analysis shows that the spectrum of the Schwarzschild mass operator is

\begin{equation}
\sigma(\hat{M})=[0,\infty).
\label{eq:continuous_mass_spectrum}
\end{equation}
Thus, within the present canonical quantization, the Schwarzschild
mass spectrum is continuous rather than discretized and the spectral resolution of the mass operator is of the form

\begin{equation}
\hat{M}=\int_{\sigma(\hat M)}d\mu(m)\,m\,\ket{m}\bra{m}.
\label{eq:continuous_mass_spectral_resolution}
\end{equation}
This result is important for the interpretation of the finite
matrices introduced below. In particular, the discrete eigenvalues
obtained after truncating the Hilbert space should not be interpreted
as a fundamental discrete mass spectrum. They represent a
finite-dimensional approximation to the underlying continuous
spectral problem.

At the end of this section, some remarks are in order. The question of whether black-hole mass has a fundamentally
discrete spectrum has a long history. In particular,
Bekenstein and Mukhanov proposed a discrete, equally spaced
horizon-area spectrum, which implies a discrete mass spectrum
for a Schwarzschild black hole \cite{BekensteinMukhanov1995}.
Their proposal is based on the assumption of fundamental
area quantization and has implications for the spectrum of
black-hole radiation. The present approach addresses a
different question: it constructs the mass operator from the
quantized Schwarzschild-interior model and investigates its
representation in a truncated Hilbert space (see the next section). Although the
mass operator has a discrete spectrum at finite truncation
dimension, this fact alone does not establish fundamental
mass quantization, since the discreteness may depend on the
truncation. In our analysis, the behavior of the spectrum
in the infinite-dimensional limit must therefore be
distinguished from the discrete area spectrum proposed by
Bekenstein and Mukhanov. The two approaches are related
through their interest in the quantum spectrum of black-hole
mass, but they employ different quantization frameworks and
need not yield the same spectral structure.

\section{Finite-dimensional spectral structure of the quantum mass observable}

Let $\left\{\ket{\Psi_n}=\ket{n,n}\right\}_{n=0}^{\infty}$,
denote the orthonormal basis of physical WDW states. To investigate the spectrum of the mass operator, we first introduce
a finite-dimensional truncation of the physical Hilbert space,

\begin{equation}
\mathcal{H}_N=\operatorname{span}\left\{\ket{\Psi_0},\ket{\Psi_1},\ldots,\ket{\Psi_{N-1}}\right\},
\label{eq:finite_physical_hilbert}
\end{equation}
where $N$ denotes the truncation dimension. The elements of the mass operator is represented in this basis by
\begin{equation}
(M_N)_{ij}=\bra{\Psi_i}\hat M\ket{\Psi_j},\qquad i,j=0,\ldots,N-1.
\label{eq:mass_matrix_definition}
\end{equation}
The corresponding truncated mass operator is then represented by the $N\times N$ matrix

\begin{equation}
M_N=\left[(M_N)_{ij}\right]_{i,j=0}^{N-1}=\begin{pmatrix}
\frac{1}{2} & -\frac{1}{2} & 0 & \cdots & 0\\
-\frac{1}{2} & \frac{3}{2} & -1 & \cdots & 0\\
0 & -1 & \frac{5}{2} & \ddots & \vdots\\
\vdots & \vdots & \ddots & \ddots & -\frac{N-1}{2}\\
0 & 0 & \cdots & -\frac{N-1}{2} & N-\frac{1}{2}
\end{pmatrix}.
\label{eq:finite_mass_matrix}
\end{equation}
Since the mass observable is Hermitian, its finite-dimensional representation satisfies $M_N^\dagger=M_N$. Consequently, all eigenvalues of $M_N$ are real and the corresponding finite-dimensional spectral problem reads

\begin{equation}
M_N \ket{v_k^{(N)}}=\lambda_k^{(N)}\ket{v_k^{(N)}},\hspace{5mm}k=0,\ldots,N-1,
\label{eq:finite_mass_eigenproblem}
\end{equation}
in which the normalized eigenvectors satisfy

\begin{equation}
\bra{v_k^{(N)}}v_l^{(N)}\rangle=\delta_{kl}.
\end{equation}
The finite-dimensional mass operator therefore admits the spectral decomposition

\begin{equation}
M_N=\sum_{k=0}^{N-1}\lambda_k^{(N)}\ket{v_k^{(N)}}\bra{v_k^{(N)}}.
\label{eq:finite_mass_spectral_decomposition}
\end{equation}
It is important, however, to distinguish the spectrum of the
truncated matrix from the spectrum of the underlying quantum
observable. The finite matrix $M_N$ is obtained by restricting
the mass operator to an $N$-dimensional subspace of the physical
Hilbert space. Therefore, its eigenvalues generally depend on the
truncation parameter $N$. In order to determine the physical spectral structure, we must
therefore investigate the behavior of $\lambda_k^{(N)}$ as the cutoff is increased.

To do this, we consider a sequence of truncated mass matrices
\begin{equation}
M_{N_1},M_{N_2},M_{N_3},\ldots,\qquad N_1<N_2<N_3<\cdots.
\end{equation}
For each value of $N$, diagonalization of $M_N$ produces a
finite set of mass eigenvalues
\begin{equation}
\lambda_0^{(N)} \leq \lambda_1^{(N)} \leq \cdots \leq \lambda_{N-1}^{(N)}.
\label{eigenvalues}
\end{equation}
A particularly useful quantity for identifying the nature of the
spectrum is the lowest eigenvalue, $\lambda_{\min}^{(N)}=\lambda_0^{(N)}$. If our numerical analysis shows that

\begin{equation}
\lambda_{\min}^{(N)} \longrightarrow 0 \qquad \text{as} \qquad N\longrightarrow \infty,
\label{eq:lowest_eigenvalue_limit}
\end{equation}
this behavior indicates that the apparent nonzero lower edge
obtained for a finite-dimensional truncation is not a fundamental
mass gap, rather, it is a truncation effect. 

The comparison of the sequences (\ref{eigenvalues}) for increasing values of $N$ allows us to distinguish two possibilities. If the eigenvalues
approach a stable discrete set, $\lambda_k^{(N)} \longrightarrow \lambda_k$, with isolated limiting values, the mass observable would possess a
discrete spectrum. On the other hand, if the eigenvalues become
increasingly dense in a continuous interval, the finite-dimensional
spectra should be interpreted as discretizations of an underlying
continuous spectrum.

It what follows we will see that the behavior observed in the present model is of the latter type.
As $N$ increases, the number of eigenvalues grows while the spacing
between neighboring eigenvalues decreases. In particular, the
spectrum accumulates toward the lower endpoint, $\lambda=0$. This suggests that the finite-dimensional spectra should not be
interpreted as evidence for a fundamental discrete mass spectrum.
Instead, they provide finite-dimensional approximations to the
spectrum of the full quantum mass operator.

The appropriate interpretation of the cutoff procedure is therefore

\begin{equation}
M_N \quad\xrightarrow[N\rightarrow\infty]{} \quad \hat M
\end{equation}
in the sense of an increasing sequence of finite-dimensional
approximations to the physical observable. Accordingly, the finite-dimensional spectral resolution (\ref{eq:finite_mass_spectral_decomposition}) should be regarded as a discretized representation of the spectral resolution of the full operator. The accumulation of eigenvalues near zero provides evidence that zero belongs to the spectrum of the full
mass operator,

\begin{equation}
0\in{\rm Spec}(\hat M).
\label{eq:zero_in_mass_spectrum}
\end{equation}
This statement does not necessarily imply the existence of a
normalizable eigenstate with exactly zero mass; rather, zero may belong
to the continuous spectrum of $\hat M$. More generally, the numerical results indicate that the physical
mass spectrum is not generated by the artificial cutoff $N$. The
cutoff merely determines the resolution with which the continuous
spectrum is represented.

This distinction will be essential in the quantum-computational
implementation developed below. A quantum computer necessarily
operates on a finite-dimensional Hilbert space, and hence a finite
matrix such as $M_N$ must be used in any practical
implementation. The parameter $N$ should therefore be interpreted as
a computational resolution parameter rather than as a physical
quantization parameter. We thus have the correspondence

\begin{eqnarray}
\text{continuous quantum observable}
\longrightarrow
\text{finite-dimensional truncation}\\ \nonumber
\longrightarrow
\text{qubit representation}.
\label{eq:continuum_to_qubits}
\end{eqnarray}
This observation provides the bridge between the quantum-gravitational
description of the Schwarzschild interior and its quantum-computing
implementation.

\section{Quantum computing}

We now construct the qubit representation of the truncated mass
observable. For this purpose, we choose the truncation dimension to
be a power of two, $N=2^n$, so that the truncated Hilbert space can be encoded into $n$ qubits,
\begin{equation}
\mathcal H_N
\simeq
(\mathbb C^2)^{\otimes n}.
\end{equation}
Under this condition the truncated mass operator becomes an $n$-qubit Hermitian
operator as represented in (\ref{eq:finite_mass_matrix}). The important conceptual distinction is that the finite-dimensional
qubit operator $\hat M_N$ is not regarded as the fundamental mass
observable. Rather, it is a computational representation of the
truncated continuous observable $\hat M$.

\subsection{Pauli decomposition of the mass operator}
\label{subsec:pauli_decomposition}

The set of all tensor products of single-qubit Pauli operators forms an orthogonal basis for operators acting on the $n$-qubit Hilbert
space. We therefore write

\begin{equation}
\hat M_N=\sum_{\alpha}c_\alpha P_\alpha,
\label{eq:mass_pauli_expansion}
\end{equation}
where

\begin{equation}
P_\alpha=\sigma_{\alpha_1}\otimes \sigma_{\alpha_2}\otimes\cdots\otimes \sigma_{\alpha_n},
\end{equation}
with

\begin{equation}
\sigma_0=I,\qquad \sigma_1=X,\qquad \sigma_2=Y,\qquad \sigma_3=Z.
\end{equation}
There are $4^n$ elements in this operator basis and their orthogonality
relation is

\begin{equation}
\operatorname{Tr}\left(P_\alpha P_\beta \right)=2^n\delta_{\alpha\beta}.
\end{equation}
Consequently, the Pauli coefficients are uniquely determined by

\begin{equation}
c_\alpha=\frac{1}{2^n}\operatorname{Tr}\left(P_\alpha\hat M_N\right).
\label{eq:pauli_coefficients}
\end{equation}
Since $\hat M_N$ is Hermitian, all coefficients $c_\alpha$ are real. The importance of the representation (\ref{eq:mass_pauli_expansion}) is that each term
$P_\alpha$ corresponds to a tensor product of elementary one-qubit
operators. Consequently, the implementation of the mass observable
on a quantum computer can be reduced to the implementation and
measurement of its Pauli strings, the number of which is

\begin{equation}
N_{\rm Pauli}=4^n=N^2.
\end{equation}
However, the actual computational complexity depends on the number
of nonvanishing coefficients,

\begin{equation}
N_{\rm nz}=\#\left\{\alpha:c_\alpha\neq0 \right\}.
\end{equation}
Thus, an important question for the quantum implementation is whether
the Pauli representation of the mass observable is sparse. We
therefore define the Pauli sparsity ratio

\begin{equation}
\mathcal S_N=\frac{N_{\rm nz}}{4^n},
\label{eq:pauli_sparsity}
\end{equation}
which measures the fraction of the complete Pauli basis that is required
to represent the mass observable.
A small value of $\mathcal S_N$ would indicate that the mass
observable admits an efficient decomposition into a relatively small
number of measurable Pauli strings, whereas a value approaching unity
would indicate a dense operator representation.

In the following, we determine the Pauli coefficients explicitly
for the finite-dimensional mass matrices obtained from the
WDW basis and investigate their scaling with the
truncation dimension $N$.

\subsubsection{Two-qubit representation}

For the smallest nontrivial qubit representation, we choose
$N=4$, corresponding to two qubits. The truncated mass matrix is

\begin{equation}
M_4=\begin{pmatrix}
\frac{1}{2} & -\frac{1}{2} & 0 & 0\\
-\frac{1}{2} & \frac{3}{2} & -1 & 0\\
0 & -1 & \frac{5}{2} & -\frac{3}{2}\\
0 & 0 & -\frac{3}{2} & \frac{7}{2}
\end{pmatrix}.
\label{eq:M4}
\end{equation}
The coefficients in the two-qubit Pauli basis are given by

\begin{equation}
c_{\alpha\beta}=\frac{1}{4}\operatorname{Tr}\left[(\sigma_\alpha\otimes\sigma_\beta)M_4\right].
\end{equation}
A direct evaluation gives seven nonvanishing coefficients,

\begin{eqnarray}
\left\{
\begin{array}{ll}
c_{II}=2,\hspace{2mm} c_{IX}=-1,\hspace{2mm}c_{IZ}=-\frac{1}{2},\\\\
c_{XX}=-\frac{1}{2},\hspace{2mm} c_{YY}=-\frac{1}{2},\hspace{2mm} c_{ZI}=-1,\\\\
c_{ZX}=\frac{1}{2}.
\end{array}
\right.
\label{eq:M4_pauli_coefficients}
\end{eqnarray}
Consequently, the two-qubit representation of the mass observable takes the form

\begin{equation}
M_4=2I\otimes I-I\otimes X-\frac{1}{2}I\otimes Z-\frac{1}{2}X\otimes X-\frac{1}{2}Y\otimes Y-Z\otimes I+\frac{1}{2}Z\otimes X.
\label{eq:M4_pauli}
\end{equation}
Thus, out of the $4^2=16$ possible two-qubit Pauli strings, only $N_{\rm nz}(4)=7$ have nonzero coefficients. The corresponding sparsity ratio is

\begin{equation}
\mathcal S_4=\frac{N_{\rm nz}(4)}{4^2}=\frac{7}{16}=0.4375.
\label{eq:sparsity_N4}
\end{equation}
The two-qubit representation also provides a direct setting for a
variational quantum estimation of the mass spectrum. In particular,
the lowest eigenvalue of $M_4$ can be obtained by minimizing the
expectation value of the Pauli-decomposed operator. We therefore
consider a variational state of the form

\begin{equation}
|\psi(\boldsymbol{\theta})\rangle=U(\boldsymbol{\theta})|00\rangle,
\end{equation}
and define the VQE cost function \cite{Tilly, Mc} as

\begin{equation}
E_M(\boldsymbol{\theta})
=
\langle\psi(\boldsymbol{\theta})|
M_4
|\psi(\boldsymbol{\theta})\rangle.
\label{eq:VQE_cost_mass}
\end{equation}
Using the Pauli decomposition in Eq.~\eqref{eq:M4_pauli}, this
expectation value can be evaluated from the corresponding Pauli
expectation values,

\begin{equation}
\begin{aligned}
E_M(\boldsymbol{\theta})={}&
2
-\langle I\otimes X\rangle
-\frac{1}{2}\langle I\otimes Z\rangle
-\frac{1}{2}\langle X\otimes X\rangle
-\frac{1}{2}\langle Y\otimes Y\rangle\\
&-\langle Z\otimes I\rangle
+\frac{1}{2}\langle Z\otimes X\rangle .
\end{aligned}
\label{eq:VQE_cost_expanded}
\end{equation}
For the present two-qubit example, we employ a shallow
hardware-efficient ansatz consisting of single-qubit $R_y$ rotations
and an entangling CNOT gate,

\begin{equation}
U(\boldsymbol{\theta})
=
\left[
R_y(\theta_3)\otimes R_y(\theta_4)
\right]
\operatorname{CNOT}
\left[
R_y(\theta_1)\otimes R_y(\theta_2)
\right].
\label{eq:two_qubit_ansatz}
\end{equation}
The variational parameters are optimized using the
L-BFGS-B classical optimizer. The calculation is performed in the
noiseless state-vector representation, so that the convergence of the
variational procedure can be compared directly with the exact
eigenvalue obtained by diagonalizing $M_4$.

\begin{figure}[t]
\centering
\includegraphics[width=0.60\textwidth]{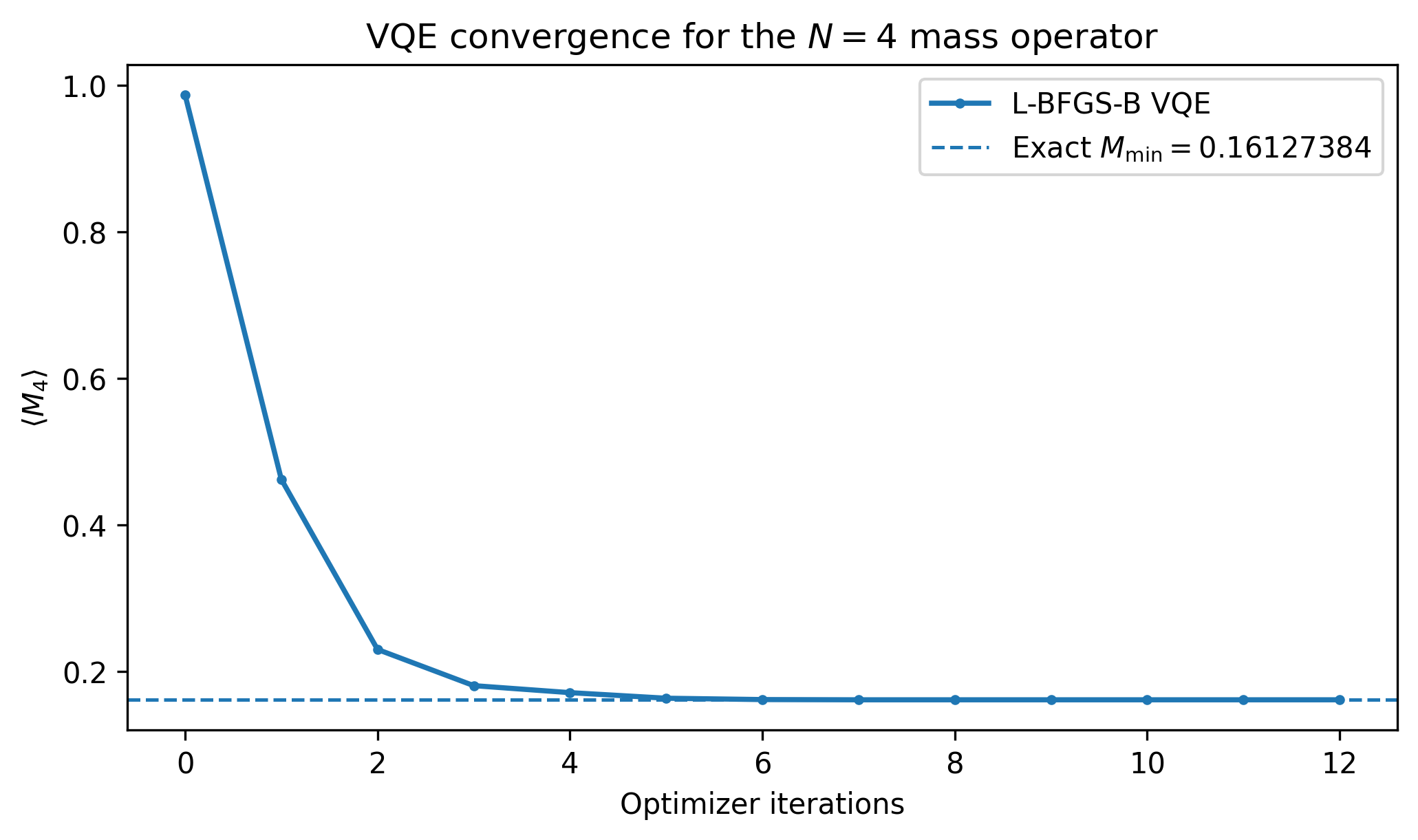}
\caption{Convergence of the VQE estimation of the lowest eigenvalue of the
$N=4$ truncated mass operator. The calculation is performed using a
two-qubit variational ansatz and the L-BFGS-B classical optimizer in
the noiseless state-vector representation. The dashed horizontal line
denotes the exact lowest eigenvalue obtained by direct diagonalization,
$M_{\min}^{(4)}=0.1612738448$.}
\label{fig:VQE_convergence_N4}
\end{figure}

The eigenvalues of the truncated mass matrix are

\begin{equation}
\operatorname{Spec}(M_4)
=\left\{0.1612738448,\,0.8728805496,\,2.2683101514,\,4.6975354542\right\},
\label{eq:M4_eigenvalues}
\end{equation}
and hence the exact ground-state value is

\begin{equation}
M_{\min}^{(4)}=0.1612738448.
\label{eq:M4_ground}
\end{equation}
The VQE optimization converges to
\begin{equation}
E_M^{\rm VQE}=0.1612738448,
\end{equation}
with an absolute deviation from the exact result at the level of
numerical precision. The convergence of the variational energy is
shown in Fig.~\ref{fig:VQE_convergence_N4}.

It is worth emphasizing that the VQE calculation is presented here
as a representative quantum benchmark rather than repeated for every
truncation size. The $N=4$ case already contains the essential
ingredients of the variational procedure: a finite-dimensional mass
operator, its Pauli decomposition, a variational ansatz, and classical
optimization of the corresponding expectation value. Increasing $N$
changes the dimension of the truncated physical Hilbert space and the
number of Pauli terms (see the following subsections), but does not introduce a qualitatively different
variational principle. The larger truncations therefore exhibit the
same type of variational convergence, while their principal new
information concerns the scaling of the qubit and measurement
complexities and the approach of the lowest mass eigenvalue toward its
large-$N$ limit. For this reason, we use the two-qubit calculation as a
representative VQE benchmark and analyze the larger values of $N$
through the exact finite-dimensional spectrum and the complexity
scaling derived above.

\subsubsection{Three-qubit representation}

We next consider the truncation $N=8$, which can be encoded into
three qubits. The corresponding mass matrix is

\begin{equation}
M_8=\begin{pmatrix}
\frac12&-\frac12&0&0&0&0&0&0\\
-\frac12&\frac32&-1&0&0&0&0&0\\
0&-1&\frac52&-\frac32&0&0&0&0\\
0&0&-\frac32&\frac72&-2&0&0&0\\
0&0&0&-2&\frac92&-\frac52&0&0\\
0&0&0&0&-\frac52&\frac{11}{2}&-3&0\\
0&0&0&0&0&-3&\frac{13}{2}&-\frac72\\
0&0&0&0&0&0&-\frac72&\frac{15}{2}
\end{pmatrix}.
\label{eq:M8}
\end{equation}
For three qubits, the Pauli expansion takes the form

\begin{equation}
M_8=\sum_{\alpha,\beta,\gamma=0}^{3}c_{\alpha\beta\gamma}\sigma_\alpha\otimes \sigma_\beta\otimes \sigma_\gamma,
\end{equation}
where

\begin{equation}
c_{\alpha\beta\gamma}=\frac{1}{8}\operatorname{Tr}\left[(\sigma_\alpha\otimes \sigma_\beta\otimes \sigma_\gamma)M_8\right].
\end{equation}
The nonvanishing coefficients are

\begin{eqnarray}
\left\{
\begin{array}{ll}
c_{III}=4,\hspace{2mm} c_{IIX}=-2,\hspace{2mm}c_{IIZ}=-\frac{1}{2},\\\\
c_{IXX}=-1,\hspace{2mm} c_{IYY}=-1,\hspace{2mm} c_{IZI}=-1,\\\\
c_{IZX}=\frac{1}{2},\hspace{2mm} c_{XXX}=-\frac{1}{2},\hspace{2mm} c_{XYY}=\frac12,\\\\
c_{YXY}=-\frac{1}{2},\hspace{2mm} c_{YYX}=-\frac{1}{2},\hspace{2mm} c_{ZII}=-2,\\\\
c_{ZIX}=1,\hspace{2mm} c_{ZXX}=\frac{1}{2},\hspace{2mm} c_{ZYY}=\frac12.
\end{array}
\right.
\label{eq:M8_pauli_coefficients}
\end{eqnarray}
Thus, the three-qubit representation of the mass observable is

\begin{equation}
\begin{aligned}
M_8={}&
4III-2IIX-\frac12IIZ-IXX-IYY-I ZI
+\frac12IZX
\\
&-\frac12XXX+\frac12XYY
-\frac12YXY-\frac12YYX
-2ZII+ZIX
\\
&+\frac12ZXX+\frac12ZYY,
\end{aligned}
\label{eq:M8_pauli}
\end{equation}
where tensor-product symbols have been suppressed for compactness. There are $4^3=64$ possible Pauli strings for three qubits, of which
only $N_{\rm nz}(8)=15$ have nonvanishing coefficients. The corresponding sparsity ratio is

\begin{equation}
\mathcal S_8=\frac{N_{\rm nz}(8)}{4^3}=\frac{15}{64}\simeq 0.2344.
\label{eq:sparsity_N8}
\end{equation}
Thus, although the absolute number of nonvanishing Pauli strings
increases with the Hilbert-space dimension, their fraction of the
complete Pauli basis decreases from $0.4375$ for $N=4$ to approximately
$0.2344$ for $N=8$.

\subsubsection{Four-qubit representation}

We next consider the truncation $N=16$, corresponding to a
four-qubit representation. The Pauli coefficients are obtained from

\begin{equation}
c_{\alpha\beta\gamma\delta}=\frac{1}{16}\operatorname{Tr}\left[(\sigma_\alpha\otimes \sigma_\beta\otimes \sigma_\gamma\otimes \sigma_\delta)M_{16}\right].
\end{equation}
A complete evaluation of the $4^4=256$ Pauli strings yields the
following nonvanishing coefficients

\begin{eqnarray}
\left\{
\begin{array}{ll}
c_{IIII}=8,\hspace{2mm} c_{IIIX}=-4,\hspace{2mm}c_{IIIZ}=-\frac{1}{2},\hspace{2mm}c_{IIXX}=-2,\hspace{2mm}c_{IIYY}=-2,\\\\
c_{IIZI}=-1,\hspace{2mm} c_{IIZX}=\frac12,\hspace{2mm}c_{IXXX}=-1,\hspace{2mm}c_{IXYY}=1,\hspace{2mm}c_{IYXY}=-1,\\\\
c_{IYYX}=-1,\hspace{2mm} c_{IZII}=-2,\hspace{2mm}c_{IZIX}=1,\hspace{2mm}c_{IZXX}=\frac12,\hspace{2mm}c_{IZYY}=\frac12,\\\\
c_{XXXX}=-\frac12,\hspace{2mm} c_{XXYY}=\frac12,\hspace{2mm}c_{XYXY}=\frac{1}{2},\hspace{2mm}c_{XYYX}=\frac12,\hspace{2mm}c_{YXXY}=-\frac12,\\\\
c_{YXYX}=-\frac12,\hspace{2mm} c_{YYXX}=-\frac12,\hspace{2mm}c_{YYYY}=\frac{1}{2},\hspace{2mm}c_{ZIII}=-4,\hspace{2mm}c_{ZIIX}=2,\\\\
c_{ZIXX}=1,\hspace{2mm} c_{ZIYY}=1,\hspace{2mm}c_{ZXXX}=\frac{1}{2},\hspace{2mm}c_{ZXYY}=-\frac12,\hspace{2mm}c_{ZYXY}=\frac12,\\\\
c_{ZYYX}=\frac12.
\end{array}
\right.
\label{eq:M16_pauli_coefficients}
\end{eqnarray}
Thus, only $N_{\rm nz}(16)=31$ out of the $256$ possible Pauli strings have nonzero coefficients.
The corresponding sparsity ratio is

\begin{equation}
\mathcal S_{16}=\frac{N_{\rm nz}(16)}{4^4}=\frac{31}{256}\simeq 0.1211.
\label{eq:sparsity_N16}
\end{equation}
The decreasing sparsity ratio indicates that the fraction of the
available Pauli basis required to represent the mass observable
decreases as the truncation dimension is increased. However, more
values of $N$ are required before a definite asymptotic scaling law
for $N_{\rm nz}$ can be established.

\subsubsection{Five-qubit representation}

For $N=32=2^5$, the mass observable is represented on five qubits.
The complete Pauli decomposition contains $4^5=1024$ possible Pauli
strings. A direct evaluation of all Pauli coefficients gives $N_{\rm nz}(32)=63$ nonvanishing terms. Hence, the corresponding sparsity ratio is

\begin{equation}
\mathcal S_{32}=\frac{N_{\rm nz}(32)}{4^5}=\frac{63}{1024}=0.0615234375.
\label{eq:sparsity_N32}
\end{equation}

\subsubsection{Six-qubit representation}

For $N=64=2^6$, the mass observable is represented on six qubits.
The complete Pauli basis contains $4^6=4096$ strings. Evaluating all
Pauli coefficients gives $N_{\rm nz}(64)=127$.
The corresponding sparsity ratio is therefore

\begin{equation}
\mathcal S_{64}=\frac{N_{\rm nz}(64)}{4^6}=\frac{127}{4096}=0.031005859375.
\label{eq:sparsity_N64}
\end{equation}

\begin{table}[ht]
\centering
\caption{Number of qubits, total number of Pauli strings, number of
nonvanishing Pauli coefficients, and the corresponding sparsity ratio
for different truncation dimensions.}
\label{tab:pauli_sparsity}
\begin{tabular}{c|c|c|c|c}
\hline
$N$ & $n_{\rm qubits}$ & $4^n=N^2$ & $N_{\rm nz}$ & $\mathcal S_N$ \\
\hline
$4$  & $2$ & $16$   & $7$   & $0.437500$ \\
$8$  & $3$ & $64$   & $15$  & $0.234375$ \\
$16$ & $4$ & $256$  & $31$  & $0.121094$ \\
$32$ & $5$ & $1024$ & $63$  & $0.061523$ \\
$64$ & $6$ & $4096$ & $127$ & $0.031006$ \\
\hline
\end{tabular}
\end{table}
The results obtained for different truncation dimensions are
summarized in Table~\ref{tab:pauli_sparsity}. Several features are apparent this Table.
First, the total number of Pauli strings grows quadratically with the
truncation dimension, $4^n=N^2$. By contrast, the number of
nonvanishing terms obtained from the explicit calculations grows
much more slowly. In particular, for the truncations considered here
we find the sequence

\begin{equation}
N_{\rm nz}=7,\;15,\;31,\;63,\;127,
\end{equation}
corresponding to

\begin{equation}
N=4,\;8,\;16,\;32,\;64.
\end{equation}
These values suggest the simple relation $N_{\rm nz}=2N-1$. Accordingly, the sparsity ratio suggested by these explicit
calculations is $\mathcal S_N=\frac{2N-1}{N^2}$. It follows that the fraction of the complete Pauli basis required
for the representation decreases as the truncation dimension is
increased. Thus, although the absolute number of nonvanishing Pauli strings
increases with $N$, the Pauli representation becomes increasingly
sparse relative to the full operator basis. In terms of the
truncation dimension, the explicit results therefore suggest a
linear growth, $N_{\rm nz}=\mathcal O(N)$, instead of the $\mathcal O(N^2)$ growth of the complete Pauli basis.

At this stage, however, these results should be regarded
as an empirical relation obtained from the finite set of truncations
considered above. In what follows, we derive this relation
analytically from the structure of the tridiagonal mass matrix.

\subsection{Scaling of the Pauli representation}
\label{subsec:pauli_scaling}

The WDW eigenstates relevant for the physical sector are
of the form $|n,n\rangle$. For notational convenience, we introduce
the shorthand $|j\rangle \equiv |j,j\rangle$, for $j=0,1,\ldots,N-1$.
In this truncated basis, the matrix elements of the mass operator are
therefore given by $(M_N)_{jk}=\langle j,j|\hat M|k,k\rangle=\langle j|\hat M|k\rangle$. Now, let $N=2^n$, so that the truncated Hilbert space is encoded into
$n$ qubits. We decompose the mass operator into its diagonal and off-diagonal parts,

\begin{equation}
\hat M_N=\hat D_N+\hat O_N.
\end{equation}
The diagonal part is

\begin{equation}
\hat D_N=\sum_{j=0}^{N-1}\left(j+\frac12\right)|j\rangle\langle j|.
\end{equation}
Writing the integer $j$ in binary form

\begin{equation}
j=\sum_{k=0}^{n-1}2^k b_k,\qquad b_k\in\{0,1\},
\label{binary}
\end{equation}
in which $b_k=\frac{1-Z_k}{2}$, where $Z_k$ denotes the Pauli-$Z$ operator acting on the $k$th qubit

\begin{equation}
Z_k|b_k\rangle=(-1)^{b_k}|b_k\rangle,
\qquad b_k\in\{0,1\},
\end{equation}
and hence its eigenvalues are $\pm1$. So we obtain

\begin{equation}
\hat D_N=\frac{N}{2}I^{\otimes n}-\frac12\sum_{k=0}^{n-1}2^k Z_k.
\label{eq:diagonal_pauli_decomposition}
\end{equation}
Thus, the diagonal part contains exactly $n+1$ nonvanishing Pauli
strings.

Also, the off-diagonal part of the mass operator is

\begin{equation}
\hat O_N=-\frac12\sum_{j=0}^{N-2}(j+1)\left(|j\rangle\langle j+1|+|j+1\rangle\langle j|\right).
\end{equation}
Introducing $k=j+1$, this can be written as

\begin{equation}
\hat O_N=-\frac12\sum_{k=1}^{N-1}k\left(|k-1\rangle\langle k|+|k\rangle\langle k-1|\right).
\label{eq:offdiagonal_mass_operator}
\end{equation}
The structure of this operator is conveniently analyzed in terms of
binary carries. Let $r$ denote the number of trailing zeros in the
binary representation of $k$. Then

\begin{equation}
k=2^r(2q+1),\qquad r=0,\ldots,n-1.
\end{equation}
The transition $k\rightarrow k-1$ changes the $r$ lowest bits from
$0$ to $1$ and the $(r+1)$-th bit from $1$ to $0$. Thus, the
off-diagonal transition associated with a fixed value of $r$ acts on
the corresponding $r+1$ qubits.

For $r=0$, the Pauli decomposition of the corresponding terms
contains $n$ distinct Pauli strings. For $r\geq1$, the transition on
the $r+1$ affected qubits produces $2^r$ distinct combinations of
$X$ and $Y$, while the dependence of the coefficient $k=2^r(2q+1)$ on the remaining $n-r$ higher-order bits produces $n-r$ independent
$I/Z$ structures. Hence, the number of nonvanishing Pauli strings
associated with a fixed $r\geq1$ is $2^r(n-r)$. The corresponding Pauli strings for different values of $r$ are
distinct, since the position of the lowest nontrivial $X/Y$ factor is
different. Therefore, the total number of nonvanishing Pauli strings
in the off-diagonal part is

\begin{align}
N_{\rm off}&=n+\sum_{r=1}^{n-1}2^r(n-r)\nonumber \\&=n+\left(2^{n+1}-2n-2\right)\nonumber \\&=2^{n+1}-n-2.,
\end{align}
Combining the diagonal and off-diagonal contributions gives

\begin{align}
N_{\rm nz}&=N_{\rm diag}+N_{\rm off}
\nonumber\\
&=(n+1)+(2^{n+1}-n-2)
\nonumber\\&=2^{n+1}-1.
\end{align}
Since $N=2^n$, we finally obtain the exact relation

\begin{equation}
N_{\rm nz}=2N-1.
\label{eq:exact_pauli_sparsity}
\end{equation}
Consequently, the sparsity ratio is

\begin{equation}
\mathcal S_N=\frac{2N-1}{N^2}.
\label{eq:exact_sparsity_ratio}
\end{equation}
In the large-$N$ limit,

\begin{equation}
\mathcal S_N \xrightarrow[N\rightarrow\infty]{}0.
\end{equation}
Thus, the Pauli representation of the mass observable becomes
increasingly sparse as the truncation dimension increases. In
particular, while the complete Pauli basis contains $N^2$ strings,
the mass operator requires only $2N-1$ nonvanishing Pauli terms.
Therefore, the number of terms scales linearly with the Hilbert-space
dimension, rather than quadratically with $N$.

\subsection{Quantum encoding of the WDW state}
\label{subsec:quantum_state_encoding}

The WDW solution obtained in the oscillator basis can be
written as (\ref{Psi}). After imposing the truncation $N=2^q$, we introduce the notation
$|j\rangle\equiv|j,j\rangle$ and obtain

\begin{equation}
|\Psi_N\rangle=\frac{1}{\sqrt{\mathcal N_N}}\sum_{j=0}^{N-1}c_j|j\rangle,
\end{equation}
where

\begin{equation}
\mathcal N_N=\sum_{j=0}^{N-1}|c_j|^2,
\end{equation}
is the normalization factor. This quantity has a direct physical and computational
interpretation: it is the fidelity between the full state and its normalized truncation

\begin{equation}
F_N=|\langle\Psi|\Psi_N\rangle|^2=\mathcal N_N.
\end{equation}
Consequently, the cutoff can be selected according to a prescribed
state-preparation accuracy,

\begin{equation}
F_N\geq 1-\epsilon.
\end{equation}
Since $N=2^q$, the truncated basis can be encoded exactly into $q$
qubits. Writing the integer $j$ in binary form (\ref{binary}), we identify

\begin{equation}
|j\rangle \longleftrightarrow |b_{q-1}\cdots b_1b_0\rangle.
\end{equation}
Therefore the WDW state is represented on the quantum register as

\begin{equation}
|\Psi_N\rangle=\sum_{j=0}^{N-1}c_j |b_{q-1}(j)\cdots b_1(j)b_0(j)\rangle.
\label{eq:qubit_encoded_state}
\end{equation}
The corresponding state-preparation problem is to construct a unitary
$U_\Psi$ satisfying

\begin{equation}
U_\Psi|0\rangle^{\otimes q}=|\Psi_N\rangle.
\label{eq:state_preparation}
\end{equation}
For real coefficients $c_n$, a recursive binary-tree construction
based on controlled $R_y$ rotations provides a direct exact
state-preparation scheme. The rotation angles are determined by
conditional probability weights of the coefficient distribution.

For a generic real $N$-dimensional state, this construction contains
$N-1$ independent rotation parameters. In the present model,
however, these parameters are not arbitrary: they are determined by
the spectral wave packet through

\begin{equation}
(m_0,\sigma_m)\longrightarrow A(m)\longrightarrow \{c_n\}\longrightarrow \{\theta_\alpha\}.
\end{equation}
Thus, the physical structure of the WDW solution directly determines the quantum state to be prepared.
Once the state has been prepared, the expectation value of the mass
observable can be evaluated using its Pauli decomposition $\hat M_N=\sum_{\ell=1}^{2N-1}\alpha_\ell P_\ell$, as

\begin{equation}
\langle\hat M_N\rangle_{\Psi_N}=\sum_{\ell=1}^{2N-1}\alpha_\ell \langle\Psi_N|P_\ell|\Psi_N\rangle.
\label{eq:mass_expectation_pauli}
\end{equation}
The linear scaling $N_{\rm nz}=2N-1$, therefore directly determines the number of
nonvanishing Pauli observables entering the measurement of the mass
expectation value.

\subsection{Commuting-group measurement of the mass observable}
\label{subsec:commuting_group_measurement}

Although the mass operator contains $2N-1$ nonvanishing Pauli
strings, these terms need not be measured independently. We now show
that they can be partitioned into a minimum number of mutually
commuting groups.

Let $N=2^n$. For each off-diagonal Pauli string, define $r$ as the
largest qubit index on which an $X$ or $Y$ operator appears. The
off-diagonal Pauli strings can then be partitioned into the groups

\begin{equation}
\mathcal G_r,\qquad r=0,1,\ldots,n-1,
\end{equation}
while all diagonal terms form an additional group
$\mathcal G_{\rm d}$.

For a fixed $r$, the Pauli strings in $\mathcal G_r$ contain only
$I$ and $Z$ operators on qubits with index larger than $r$, an $X$ or
$Y$ operator on qubit $r$, and an even parity of $Y$ operators on the
lower-order qubits. Consequently, any two Pauli strings belonging to
the same group $\mathcal G_r$ commute. The diagonal group is also
mutually commuting since it contains only $I$ and $Z$ operators.
Thus, this construction provides a commuting partition containing
$n+1$ groups, and hence

\begin{equation}
N_{\rm settings}\leq n+1.
\end{equation}
To establish that this bound is optimal, consider the Pauli strings

\begin{equation}
A_0=X_0,
\end{equation}
and

\begin{equation}
A_r=Y_r X_{r-1}\cdots X_1Y_0,
\end{equation}
for $r\geq1$. Each $A_r$ belongs to the corresponding group $\mathcal G_r$.
Moreover, for any $r<s$, the operators $A_r$ and $A_s$ anticommute,
because at qubit $r$ they contain $Y_r$ and $X_r$, respectively,
while all remaining anticommutations occur in pairs. Hence,

\begin{equation}
\{A_0,A_1,\ldots,A_{n-1}\}
\end{equation}
is a set of pairwise anticommuting Pauli strings.

The diagonal Pauli operator $Z_0$ anticommutes with every $A_r$,
since all $A_r$ contain either $X_0$ or $Y_0$. Therefore,

\begin{equation}
\left\{Z_0,A_0,A_1,\ldots,A_{n-1}
\right\}
\end{equation}
contains $n+1$ pairwise anticommuting Pauli strings. Since two
anticommuting Pauli operators cannot belong to the same commuting
measurement group, any measurement partition requires at least
$n+1$ groups:

\begin{equation}
N_{\rm settings}\geq n+1.
\end{equation}
Combining the upper and lower bounds gives the exact result

\begin{equation}
N_{\rm settings}=n+1.
\end{equation}
Since $N=2^n$, the number of independent measurement settings can be written as

\begin{equation}
N_{\rm settings}=\log_2N+1.
\label{eq:minimum_measurement_settings}
\end{equation}
Thus, although the number of nonvanishing Pauli terms scales
linearly with the Hilbert-space dimension, $N_{\rm nz}=\mathcal O(N)$, the minimum number of mutually commuting measurement settings grows
only logarithmically, $N_{\rm settings}=\mathcal O(\log N)$.

\subsection{Truncation accuracy and qubit requirements}
\label{subsec:truncation_accuracy}

The truncation dimension required for a reliable quantum simulation
depends on the parameters of the spectral wave packet. We quantify
this dependence through the truncation fidelity

\begin{equation}
F_N(m_0,\sigma_m)=\sum_{n=0}^{N-1}|c_n|^2,
\end{equation}
where the coefficients $c_n$ are determined by the spectral
amplitude $A(m)$. For a Hilbert-space dimension
$N=2^q$, the minimum number of qubits required to achieve a target
fidelity $1-\epsilon$, can therefore be defined as

\begin{equation}
q_{\min}(m_0,\sigma_m,\epsilon)=\min_q\left\{q:F_{2^q}(m_0,\sigma_m)\geq1-\epsilon \right\}.
\end{equation}
As a representative numerical study, we evaluated the truncation
fidelity for Gaussian spectral wave packets with several values of
$m_0$ and $\sigma_m$. The resulting minimum truncation dimensions
for $F_N\geq0.99$ and $F_N\geq0.999$ are summarized in
Table~\ref{tab:truncation_requirements}.

\begin{table}[ht]
\centering
\caption{Minimum Hilbert-space dimension and number of qubits required
to achieve target truncation fidelities for Gaussian spectral wave
packets. Here $N=2^q$.}
\label{tab:truncation_requirements}
\begin{tabular}{cccccc}
\toprule
$m_0$ & $\sigma_m$ &
\multicolumn{2}{c}{$F_N\geq0.99$} &
\multicolumn{2}{c}{$F_N\geq0.999$} \\
\cmidrule(lr){3-4}
\cmidrule(lr){5-6}
&
&
$N_{\min}$ & $q_{\min}$ &
$N_{\min}$ & $q_{\min}$ \\
\midrule
0.5 & 0.1 & 64  & 6 & 128 & 7 \\
0.5 & 0.2 & 16  & 4 & 32  & 5 \\
0.5 & 0.3 & 8   & 3 & 16  & 4 \\
0.5 & 0.5 & 4   & 2 & 8   & 3 \\
1.0 & 0.1 & 128 & 7 & 256 & 8 \\
1.0 & 0.2 & 32  & 5 & 64  & 6 \\
1.0 & 0.3 & 16  & 4 & 16  & 4 \\
1.0 & 0.5 & 4   & 2 & 8   & 3 \\
2.0 & 0.1 & 256 & 8 & 512 & 9 \\
2.0 & 0.2 & 64  & 6 & 128 & 7 \\
2.0 & 0.3 & 32  & 5 & 64  & 6 \\
2.0 & 0.5 & 8   & 3 & 16  & 4 \\
\bottomrule
\end{tabular}
\end{table}
The results demonstrate that the required truncation dimension is
strongly dependent on the parameters of the spectral wave packet.
For fixed $\sigma_m$, increasing the central mass $m_0$ shifts
the wave packet toward higher oscillator levels and consequently
increases the required Hilbert-space dimension. Conversely, broader
spectral wave packets can be represented with substantially smaller
values of $N$. The number of qubits follows directly from
$q=\log_2N$.

It is also noteworthy that increasing the target fidelity from
$0.99$ to $0.999$ can require an additional qubit. This is a
consequence of the binary encoding, for which increasing the number
of qubits by one doubles the available Hilbert-space dimension.
For example, for $m_0=1$ and $\sigma_m=0.2$, the fidelities at
$N=32$ and $N=64$ are approximately $0.9975$ and $0.99997$,
respectively. Thus, $N=32$ is sufficient for a $99\%$ fidelity,
whereas $N=64$ is required to reach $99.9\%$ fidelity.

The advantage of measurement grouping becomes increasingly
significant as the truncation dimension grows. In the absence of
grouping, one would associate one independent measurement basis with
each Pauli string, resulting in a cost proportional to
$2N-1$. With the optimal grouping obtained above, the corresponding
number of independent settings is only $\log_2N+1$. The reduction
factor in the number of settings is therefore

\begin{equation}
\mathcal R_N=\frac{2N-1}{\log_2N+1},
\end{equation}
which asymptotically behaves as

\begin{equation}
\mathcal R_N \sim \frac{2N}{\log_2N}.
\end{equation}
The overall resource requirements can consequently be summarized by the chain

\begin{equation}
(m_0,\sigma_m,\epsilon)\longrightarrow q_{\min}\longrightarrow \left(N_P,\,N_{\rm settings}\right),
\end{equation}
with
\begin{equation}
N_P=2^{q_{\min}+1}-1,\qquad N_{\rm settings}=q_{\min}+1.
\end{equation}
This establishes a direct connection between the physical
parameters of the spectral mass wave packet, the accuracy of the
Hilbert-space truncation, and the quantum resources required for
representing and measuring the mass observable.

The finite-dimensional Hilbert-space representation used here
should be understood as a controlled truncation of the underlying
quantum model, rather than as an assumption that the black-hole
interior possesses a fundamentally finite-dimensional state space.
Its physical relevance therefore depends on whether the quantities
of interest exhibit stable behavior as the truncation dimension is
increased. In the present framework, the finite-dimensional
representations provide a systematic means of investigating the
mass operator, its expectation values and fluctuations, and the
convergence of its spectral properties toward the continuum
description. This distinction is important: the finite-dimensional
model is a computational approximation to the quantum dynamics
defined by the underlying constraint, and its physical predictions
must be assessed through convergence and consistency checks.

Although the present analysis does not describe black-hole
evaporation or derive the thermodynamic entropy of a black hole,
it may provide a useful starting point for further investigations
of these questions. For example, once an appropriate physical
state and subsystem decomposition have been specified, the
quantum states considered in this framework could be used to
investigate quantum correlations and entanglement-based measures
of entropy. Addressing black-hole evaporation, however, would
require additional ingredients, such as the coupling to exterior
degrees of freedom or quantum fields and a treatment of the
relevant dynamical evolution. These extensions lie beyond the
scope of the present work.

\section{Verification of the WDW constraint}
\label{subsec:constraint_verification}

The Hamiltonian constraint admits a particularly simple representation
in the two-mode oscillator basis. With the oscillator convention used
in the WDW analysis, one has

\begin{equation}
\hat p_u^2+\frac14\hat u^2=\hat n_u+\frac12,\qquad \hat p_v^2+\frac14\hat v^2=\hat n_v+\frac12,
\end{equation}
and hence

\begin{equation}
\hat{\mathcal H}=\hat n_u-\hat n_v.
\label{Hamil}
\end{equation}
Therefore, in the complete two-mode basis $\{|n,m\rangle\}$,

\begin{equation}
\hat{\mathcal H}|n,m\rangle=(n-m)|n,m\rangle.
\end{equation}
For a truncation in which each oscillator is restricted to
$N$ states, the corresponding Hilbert space has dimension $N^2$,
and the Hamiltonian constraint is represented by the diagonal matrix

\begin{equation}
H_N=\operatorname{diag}\left\{n-m\right\}_{n,m=0}^{N-1}.
\end{equation}
The physical sector considered in the present model is spanned by
the states $|n,n\rangle$. Consequently,

\begin{equation}
\hat{\mathcal H}|n,n\rangle=0,
\end{equation}
and the WDW constraint is identically satisfied on this
subspace.

For quantum verification of the constraint, it is more informative
to consider the positive semidefinite operator $\hat{\mathcal H}^2$.
For a general state

\begin{equation}
|\Psi\rangle=\sum_{n,m}c_{nm}|n,m\rangle,
\end{equation}
one obtains

\begin{equation}
\langle\hat{\mathcal H}^2\rangle=\sum_{n,m}(n-m)^2|c_{nm}|^2.
\end{equation}
It follows that

\begin{equation}
\langle\hat{\mathcal H}^2\rangle=0
\quad\Longleftrightarrow\quad
\hat{\mathcal H}|\Psi\rangle=0.
\end{equation}
Thus, $\langle\hat{\mathcal H}^2\rangle$ provides a direct quantitative
measure of constraint violation. In contrast to
$\langle\hat{\mathcal H}\rangle$, which may vanish through cancellations
between positive and negative contributions, the expectation value of
$\hat{\mathcal H}^2$ vanishes only for states entirely contained in
the physical subspace.

Since both $\hat{\mathcal H}$ and $\hat{\mathcal H}^2$ are diagonal in
the computational oscillator basis, their expectation values can be
obtained using a single measurement basis. This provides a particularly
simple quantum-computational test of the WDW constraint. Let each oscillator be encoded
using $q$ qubits, with $N=2^q$. A computational basis state of the
two-mode system is then represented as

\begin{equation}
|n,m\rangle=|n\rangle_u\otimes|m\rangle_v,
\end{equation}
where

\begin{equation}
n=\sum_{k=0}^{q-1}2^k b_k^{(u)},\qquad m=\sum_{k=0}^{q-1}2^k b_k^{(v)}.
\end{equation}
According to (\ref{Hamil}), a measurement in the computational basis directly determines the
eigenvalue of the constraint for each measurement shot,

\begin{equation}
h=n-m.
\end{equation}
The expectation value of the positive operator $\hat{\mathcal H}^2$
can therefore be estimated without an ancilla or an additional
quantum subroutine as

\begin{equation}
\langle\hat{\mathcal H}^2\rangle\simeq \frac{1}{N_{\rm shots}}\sum_{s=1}^{N_{\rm shots}}(n_s-m_s)^2.
\end{equation}
For an ideal physical state

\begin{equation}
|\Psi_{\rm phys}\rangle=\sum_{n=0}^{N-1}c_n|n,n\rangle,
\label{IdealState}
\end{equation}
all measurement outcomes satisfy $n=m$, and hence

\begin{equation}
\langle\hat{\mathcal H}^2\rangle=0.
\end{equation}
Any outcome with $n\neq m$ signals leakage outside the physical
constraint sector. The corresponding leakage probability is

\begin{equation}
P_{\rm leak}=\sum_{n\neq m}|c_{nm}|^2,
\end{equation}
whereas

\begin{equation}
\langle\hat{\mathcal H}^2\rangle=\sum_{n,m}(n-m)^2|c_{nm}|^2,
\end{equation}
also quantifies the magnitude of the constraint violation.

Thus, the WDW constraint can be verified using a single
computational measurement basis. The procedure requires $2q$ qubits
for the two oscillator modes and only one independent measurement
setting,

\begin{equation}
N_{\rm qubits}=2q,
\qquad
N_{\rm settings}^{(\mathcal H^2)}=1.
\end{equation}

\subsection{A four-qubit demonstration}
\label{subsec:four_qubit_demo}

\begin{figure}[t]
\centering
\includegraphics[width=0.85\textwidth]{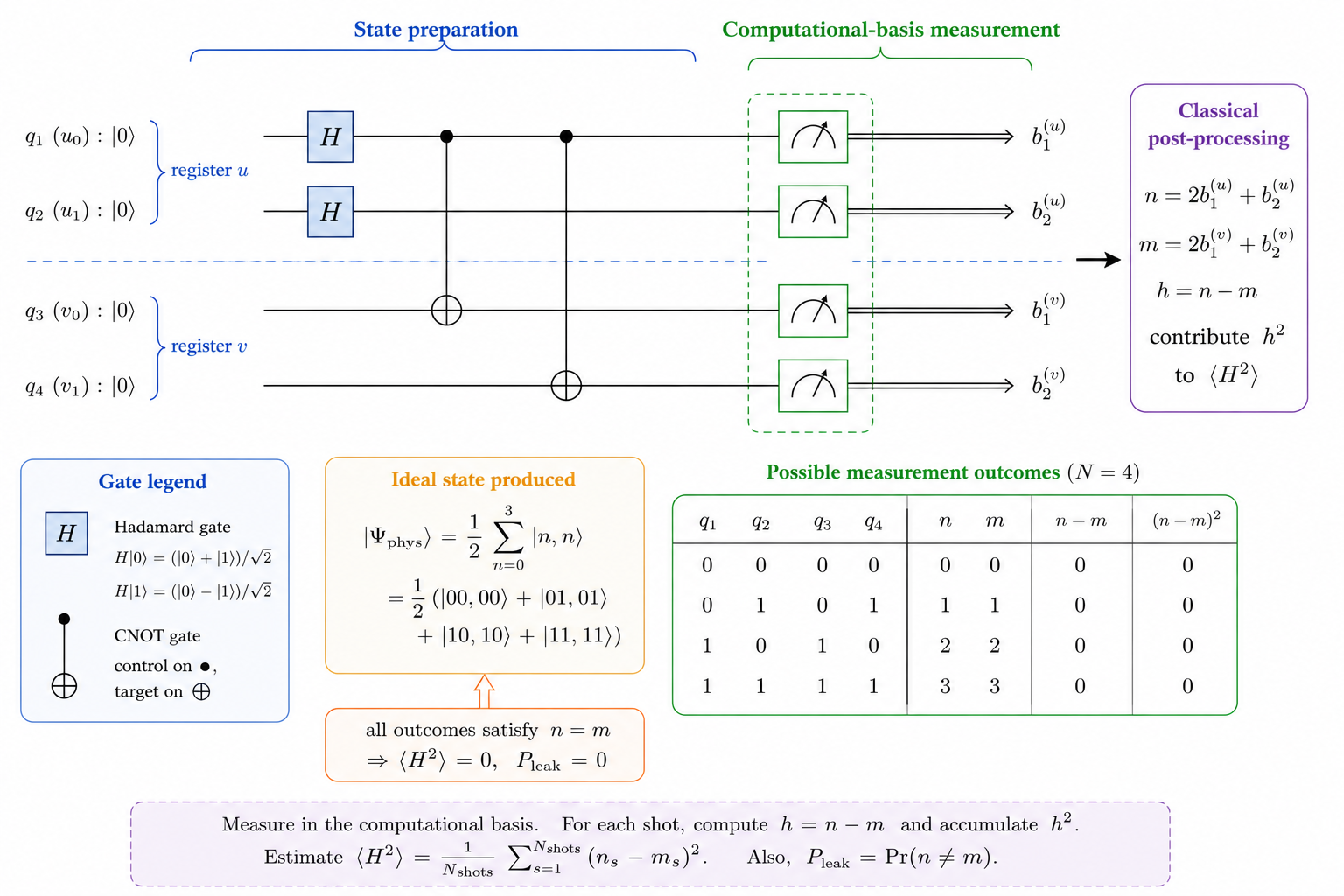}
\caption{Four-qubit circuit for preparing and verifying a physical state in the
$N=4$ truncated model. Hadamard gates create an equal superposition
in the $u$ register, while the two CNOT gates correlate the two
registers and produce
$\left|\Psi_{\rm phys}\right\rangle
=\frac{1}{2}\sum_{n=0}^{3}|n,n\rangle$.
A computational-basis measurement of the four qubits verifies the
correlation $n=m$ and therefore the WDW constraint.}
\label{fig:constraint_circuit_N4}
\end{figure}

As a simple demonstration of the constraint-verification procedure,
consider the smallest nontrivial truncation with $N=4=2^2$. Each oscillator is then encoded in two qubits, so that the complete
two-mode system requires four qubits. As illustrated in Fig.~\ref{fig:constraint_circuit_N4}, starting from the computational
basis state $|00\rangle_u|00\rangle_v$, two Hadamard
gates first generate an equal superposition of the computational
states of the first oscillator. Two CNOT gates then correlate the
corresponding qubits of the two oscillator registers. The resulting
state is

\begin{equation}
|\Psi_{\rm phys}\rangle=\frac{1}{2}\sum_{n=0}^{3}|n,n\rangle.
\label{4qstate}
\end{equation}
All computational-basis measurement outcomes therefore satisfy
$n=m$, and consequently

\begin{equation}
P_{\rm leak}=0,
\qquad
\langle\hat{\mathcal H}^2\rangle=0.
\end{equation}
The same circuit also provides a direct test of constraint violation.
For example, consider a state containing a small leakage component,
\begin{equation}
|\Psi_\epsilon\rangle=\sqrt{1-\epsilon}\,|\Psi_{\rm phys}\rangle+\sqrt{\epsilon}\,|0,1\rangle.
\end{equation}
In this case, $P_{\rm leak}=\epsilon$, while $\langle\hat{\mathcal H}^2\rangle=\epsilon$. More generally, leakage into a state
$|n,m\rangle$ contributes $(n-m)^2$ to the constraint-violation
measure. Thus, $P_{\rm leak}$ quantifies the probability of leaving
the physical sector, whereas $\langle\hat{\mathcal H}^2\rangle$
also accounts for the magnitude of the violation.

This example illustrates that constraint verification requires only
the computational measurement basis and does not require an additional quantum subroutine or ancilla. For the two-mode $N=4$
truncation, the procedure uses four qubits and a single independent
measurement setting.

\subsection{Mass estimation in the physical sector}
\label{subsec:mass_estimation}

Once the WDW constraint has been verified, the same
quantum state can be used to extract the physical mass observable.
For a truncated physical state (\ref{IdealState}), the expectation value and variance of the mass are given by

\begin{equation}
\langle M\rangle=\langle\Psi_{\rm phys}|\hat M_N|\Psi_{\rm phys}\rangle,\qquad (\Delta M)^2=\langle M^2\rangle-\langle M\rangle^2.
\end{equation}
Using the Pauli decomposition of the truncated mass operator (\ref{eq:mass_pauli_expansion}), the expectation value can be evaluated as

\begin{equation}
\langle M\rangle=\sum_{\alpha}c_{\alpha}\langle P_{\alpha}\rangle.
\end{equation}
The number of independent measurement settings is therefore the same
as (\ref{eq:minimum_measurement_settings}), that obtained previously. As a simple illustration, consider the $N=4$ physical state (\ref{4qstate}). Using the corresponding truncated mass matrix, one obtains $\langle M\rangle=\frac{1}{2}$, and $\langle M^2\rangle=1$, and hence $(\Delta M)^2=\frac{3}{4}$.
At the same time, the state exactly satisfies the WDW
constraint, $\langle\hat{\mathcal H}^2\rangle=0$. Thus, constraint satisfaction does not require the state to be a mass
eigenstate: a physical state may contain a nonzero mass uncertainty.

\section{Conclusions}
\label{sec:conclusions}

In this work, we have developed a quantum-computational framework
for the mass observable of a Schwarzschild black-hole interior,
starting from its WDW description. Working in the
oscillator basis of the WDW equation, the physical sector
is identified with the subspace spanned by the two-mode states
$|n,n\rangle$. This structure provides a natural starting point for
constructing a finite-dimensional quantum representation of the
black-hole mass.

We first truncated the mass operator to an $N$-dimensional physical
subspace and studied its finite-dimensional spectrum. The numerical
analysis for increasing truncation sizes,
$N=4,8,16,32,$ and $64$, shows a systematic decrease of the lowest
eigenvalue and supports the approach toward a continuous mass
spectrum in the large-$N$ limit. The truncation therefore provides a
controlled route from the infinite-dimensional WDW
description to a finite-dimensional quantum representation suitable
for quantum computation. Beyond its computational role, the truncated representation
provides a framework for systematic studies of quantum
observables and correlations in the Schwarzschild interior,
while its possible implications for black-hole entropy and
evaporation require further analysis incorporating additional
physical degrees of freedom and dynamics.

A central result of the present analysis is the explicit Pauli
decomposition of the truncated mass operator. For the $N=4$ case,
the operator can be represented using two qubits, and only seven of
the sixteen possible Pauli strings have nonzero coefficients. More
generally, we established the scaling $N_{\rm nz}=2N-1$, for the number of nonvanishing Pauli terms. This result shows that,
although the Hilbert-space dimension grows as $N$, the number of
terms required in the Pauli representation grows only linearly with
the truncation size. The corresponding sparsity decreases as $\mathcal{S}_N=\frac{2N-1}{N^2}$, for $N$ restricted to powers of two.

We have also analyzed the measurement complexity associated with
this decomposition. By grouping mutually compatible Pauli operators,
the number of independent measurement settings is found to scale as $N_{\rm settings}=\log_2N+1$. Thus, the number of qubits required to encode the truncated physical
mass operator grows logarithmically with the truncation dimension,
while the number of nonzero Pauli terms grows linearly and the number
of independent measurement settings grows logarithmically. These
scalings provide a direct characterization of the quantum
complexity of the mass observable.

As a representative quantum benchmark, we implemented a
two-qubit variational quantum eigensolver for the $N=4$ mass operator.
The VQE minimizes the expectation value of the Pauli-decomposed mass
operator and reproduces the lowest eigenvalue obtained by exact
diagonalization. This calculation demonstrates explicitly that the
finite-dimensional mass representation can be translated into a
variational quantum computation. The VQE calculation was used as a
representative benchmark rather than repeated for larger truncations,
since increasing $N$ changes the size and complexity of the
representation but does not introduce a qualitatively different
variational procedure.

Finally, we discussed the verification of the WDW
constraint independently of the mass estimation. In the two-mode
representation, computational-basis measurements provide a direct
test of the correlation $n=m$, or equivalently of
$\langle\hat{\mathcal H}^2\rangle=0$ for an ideal physical state.
Once the constraint is satisfied, the mass expectation value and its
fluctuations can be extracted from the Pauli representation of the
mass operator. This separates two conceptually distinct tasks:
verifying that a state belongs to the physical WDW sector
and extracting physical information from that state.

The results establish a concrete bridge between quantum cosmological
black-hole models and near-term quantum-computational techniques.
Rather than treating the quantum computer merely as a numerical
diagonalization device, our construction identifies the resources
required to encode, measure, verify, and variationally analyze a
physical observable of the black-hole interior. The logarithmic
qubit and measurement scaling, together with the linear growth of the
Pauli representation, suggests that the framework remains
computationally structured as the truncation is increased.

Several directions remain open. A natural extension is to study
larger-scale variational calculations in the presence of realistic
sampling noise and gate errors, as well as alternative ans\"atze
optimized for the structure of the mass operator. It would also be
interesting to investigate other black-hole observables and other
quantum-gravitational minisuperspace models within the same
matrix-to-Pauli framework. These extensions may provide a systematic
route toward testing quantum-gravitational observables on
programmable quantum processors.

\end{document}